# CAUSALLY SYMMETRIC BOHM MODEL

**Roderick I. Sutherland**

Centre for Time, Department of Philosophy, University of Sydney, NSW 2006 Australia

## Abstract

A version of Bohm's model incorporating retrocausality is presented, the aim being to explain the nonlocality of Bell's theorem while maintaining Lorentz invariance in the underlying ontology. The strengths and weaknesses of this alternative model are compared with those of the standard Bohm model.

## Article outline

## 1. Introduction

The aim of this paper is to construct a version of Bohm's model that also includes the existence of backwards-in-time influences in addition to the usual forwards causation. The motivation for this extension is to remove the need in the existing model for a preferred reference frame. As is well known, Bohm's explanation for the nonlocality of Bell's theorem necessarily involves instantaneous changes being produced at space-like separations, in conflict with the "spirit" of special relativity even though these changes are not directly observable. While this mechanism is quite adequate from a purely empirical perspective, the overwhelming experimental success of special relativity (together with the theory's natural attractiveness), makes one reluctant to abandon it even at a "hidden" level. There are, of course, trade-offs to be made in formulating an alternative model and it is ultimately a matter of taste as to which is preferred. However, constructing an explicit example of a causally symmetric formalism allows the pros and cons of each version to be compared and highlights the consequences of imposing such symmetry[1]. In particular, in addition to providing a natural explanation for Bell nonlocality, the new model allows us to define and work with a mathematical description in 3-dimensional space, rather than configuration space, even in the correlated many-particle case.

The structure of the paper is as follows. In section 2, the basic causally symmetric scheme is introduced in terms of initial and final boundary conditions. Section 3 then highlights the ways in which the corresponding initial and final wavefunctions will propagate. The basic equations of the alternative model are deduced in section 4 in close analogy to the formalism of the standard Bohm model. Section 5 then points out how the notion of retrocausality has been given an explicit mathematical form and section 6 checks some elementary matters of consistency. The discussion in section 7 indicates how backwards-in-time effects provide a meaning for the notion of negative probability. Section 8 then explains the way in which a possible objection to the model is overcome. After dealing with some technical details in section 9, the analysis in section 10 shows how the model explains Bell's nonlocality in a way that is Lorentz invariant, as well as being local from a 4-dimensional point of view. The generalization of the formalism to many particles is given in sections 11 and 12. A theory of measurement is outlined in section 13 for comparison with that of the standard Bohm model, then a relativistic version of the causally symmetric approach is formulated in section 14 for the single-particle Dirac case. Finally, conclusions are presented in section 15.

## 2. General structure of the model

We will limit ourselves initially to the single-particle case for simplicity. The many-particle case will be considered later after gaining some preliminary insight from a discussion of the EPR/Bell arrangement.

---

[1] This notion of causal symmetry needs to be distinguished from the more usual concept of time symmetry. Most mathematical formalisms in physics, including the Bohm model, already possess symmetry under time reversal, but this is separate from the issue of causal structure.

Bohm's model[2] makes the assumption that a particle always has a definite, but hidden, trajectory. It then specifies the particle's velocity in terms of the wavefunction $\psi$. Our aim here is to provide a consistent generalization of this formalism that incorporates backwards-in-time effects, or retrocausality, into the model. The state of the particle at any time will then be partly determined by the particle's future experiences as well as by its past.

As a first step towards developing such a formalism, we must deal with the question: what aspects of a particle's future are relevant?[3] Possible factors could be the type of measurement to be performed next, the nature of the particle's interaction with the next particle it encounters and perhaps the nature of all future measurements and interactions. This seems a daunting prospect at first. However, an indication of the best way to proceed is obtained by looking at the usual way we take account of a particle's **past** experiences: we work with an initial wavefunction $\psi_i$ which summarizes the particle's relevant past. More formally speaking, $\psi_i$ specifies the initial boundary conditions. Therefore, by symmetry, it seems natural to supplement $\psi_i$ with a "final" wavefunction $\psi_f$ specifying the **final** boundary conditions. To keep the arrangement time-symmetric, the final wavefunction $\psi_f$ will be restricted, like $\psi_i$, to being a solution of the time-dependent Schrödinger equation. The procedure to be followed here then is to construct a version of Bohm's model containing both $\psi_i$ and $\psi_f$.

Note that the new wavefunction $\psi_f$ being introduced here is independent of the usual wavefunction $\psi_i$ and should not be confused with the result of evolving $\psi_i$ deterministically to a later time. Thus, at any single time t, there are two distinct wavefunctions: (i) the initial wavefunction $\psi_i(\mathbf{x},t)$, which summarizes the initial boundary conditions existing at some earlier time $t_1$ and which has been evolved forwards from $t_1$ to t and (ii) the final wavefunction $\psi_f(\mathbf{x},t)$, which summarizes the final boundary conditions at some later time $t_2$ and which has been evolved back from $t_2$ to t. The model to be developed here will be deterministic once **both** wavefunctions are specified, together with the particle's position at one instant of time. In particular, specifying $\psi_i$ at time $t_1$ and $\psi_f$ at time $t_2$ will then determine the particle's velocity at any intermediate time.

Like the standard Bohm model, the causally symmetric version will be a "no collapse" model, with empty branches of wavefunctions after measurements being ignored as irrelevant. The model does not give any special status to measurement interactions, observers or the macroscopic world. Indeed, it is intended to be as similar as possible to the standard Bohm formulation, apart from the obvious fact that such a retrocausal model cannot be deterministic when only the initial conditions are given.

---

[2] Bohm, 1952a, 1952b.

[3] Some of the presentation in this paper has been employed previously in Sutherland (1998).

## 3. Backwards-in-time effects

At first sight, it may seem that the model being proposed is simply one containing a second wavefunction (acting as a hidden variable) without necessarily being retrocausal[4]. It is important, therefore, to note certain ways in which $\psi_i$ and $\psi_f$ differ in behaviour. For example:

(i) Consider a particle propagating from a source to a photographic plate. Its $\psi_i$ typically spreads out forwards in time in propagating from the source to the plate. By contrast, the particle's $\psi_f$ typically spreads out backwards in time in going from the spot on the photographic plate back to the source.

(ii) Consider a particle which is initially isolated but which then interacts with other particles before eventually being detected. Starting as a single-particle wavefunction, the particle's $\psi_i$ will evolve forward in time to form a correlated, many-particle wavefunction in 3n-dimensional configuration space. By contrast, the particle's $\psi_f$ will be a single-particle wavefunction at the final detection point, with the interactions making it more and more correlated in going backwards in time towards the source.

The model's retrocausal nature is highlighted further in section 5.

## 4. Basic mathematical formalism

The standard version of Bohm's model will now be summarized briefly for comparison with the equations of the subsequent causally symmetric model. Strictly speaking the wavefunctions in this summary should all be written with subscripts i for "initial", to conform with the notation introduced above. For simplicity, however, the i's will not be included here.

For the single-particle case we are initially considering, Bohm's model postulates the following:

(i) For a particle with wavefunction $\psi(\mathbf{x},t)$, the probability distribution $\rho(\mathbf{x},t)$ for the position $\mathbf{x}$ of the particle at any time t is given by

$$\rho(\mathbf{x},t) = \psi^*\psi \tag{1}$$

(ii) The velocity $\mathbf{v}(\mathbf{x},t)$ of the particle is related at all times to the particle's position by

$$\mathbf{v} \equiv \frac{d\mathbf{x}}{dt} = \frac{\hbar}{2im}\frac{\psi^*\overleftrightarrow{\nabla}\psi}{\psi^*\psi} \tag{2}$$

---

[4] It is, of course, possible to construct a two-wavefunction model that is not causally symmetric and it may be possible to construct a single-wavefunction model containing retrocausality. Neither, however, is relevant to the present aim of formulating a Lorentz invariant Bohm model.

where m is the particle's mass, $\hbar$ is Planck's constant, $\overleftrightarrow{\nabla}$ stands for $\overrightarrow{\nabla} - \overleftarrow{\nabla}$ and the grad operators $\overrightarrow{\nabla}$ and $\overleftarrow{\nabla}$ act to the right and left, respectively. As shown by Bohm, the model characterized by these assumptions is consistent with all the predictions of non-relativistic quantum mechanics. Given the initial position of a particle, equation (2) uniquely determines the particle's future trajectory and so the above scheme is deterministic. The need to resort to the usual statistical description of quantum mechanics is then attributed in this model to our inherent lack of knowledge of the particle's initial position within the wavefunction.

The notation is usually simplified by writing the wavefunction in the polar form:

$$\psi = R\, e^{iS/\hbar} \tag{3}$$

where $R(\mathbf{x}, t)$ and $S(\mathbf{x}, t)$ are real quantities. Equations (1) and (2) may then be expressed as:

$$\rho(\mathbf{x}, t) = R^2 \tag{4}$$

and

$$\mathbf{v} = \frac{\nabla S}{m} \tag{5}$$

This simplification, however, is not always available in relativistic versions of Bohm's model. In particular, it is not possible in Bohm's model for the Dirac equation[5]. In the present context, the polar notation of (3), (4) and (5) does not provide any obvious advantage and so will not be employed.

There are several arguments that lead to the choice of velocity expression in equation (2). The most useful one for our present purposes will now be outlined to serve as a basis for obtaining a causally symmetric version. In the standard formalism for the flow of probability current, the evolution of probability density with time is analogous to the flow of a fluid. In order for the probability to be conserved at each point, it must satisfy the equation of continuity:

$$\nabla \cdot (\rho \mathbf{v}) + \frac{\partial \rho}{\partial t} = 0 \tag{6}$$

where $\rho(\mathbf{x}, t)$ is the probability density for the particle to be within a volume element $d^3\mathbf{x}$ surrounding position $\mathbf{x}$ at time t, and $\mathbf{v}(\mathbf{x}, t)$ is the velocity of the probability flow at that point. Bohm's model involves the extra assumption that there is a unique particle velocity specified at each point in space-time once the wavefunction is given. In these circumstances, the velocity of the probability flow at $(\mathbf{x}, t)$ is the same as the particle velocity at that point. Therefore, in constructing Bohm's model, the expressions chosen in terms of $\psi$ for the particle's velocity $\mathbf{v}(\mathbf{x}, t)$ and position probability density $\rho(\mathbf{x}, t)$ must together satisfy equation (6) in order to conserve probability at each point[6].

---

[5] Bohm, 1953.

[6] Local conservation of probability here essentially means compatibility with the existence of unbroken trajectories, so that particles need not be spontaneously appearing and disappearing.

Now, starting from the Schrödinger equation:

$$-\frac{\hbar^2}{2m}\nabla^2\psi + V\psi = i\hbar\frac{\partial\psi}{\partial t} \tag{7}$$

and its complex conjugate:

$$-\frac{\hbar^2}{2m}\nabla^2\psi^* + V\psi^* = -i\hbar\frac{\partial\psi^*}{\partial t} \tag{8}$$

the corresponding equation of continuity resembling (6) is obtained by the familiar method of multiplying (7) by $\psi^*$ and (8) by $\psi$, then subtracting the resulting two equations, to obtain:

$$\nabla\cdot\left(\frac{\hbar}{2im}\psi^*\overleftrightarrow{\nabla}\psi\right) + \frac{\partial}{\partial t}(\psi^*\psi) = 0 \tag{9}$$

This equation holds automatically for any wavefunction satisfying the Schrödinger equation. Comparing (6) and (9) then points to the expressions chosen in equations (1) and (2), so that Bohm's model is thereby obtained.

The aim now is to follow an analogous path to a causally symmetric version of Bohm's model. Such a model must obviously feature both the initial and final wavefunctions $\psi_i$ and $\psi_f$ in its formalism. The key point to note is that the steps leading to equation (9) essentially treat $\psi$ and $\psi^*$ as two separate functions and do not depend on them being related as complex conjugates. Indeed, if one takes them as independent functions by simply putting a subscript i on $\psi$ and a subscript f on $\psi^*$, the Schrödinger equation ensures that the following modified version of equation (9) still holds:

$$\nabla\cdot\left(\frac{\hbar}{2im}\psi_f^*\overleftrightarrow{\nabla}\psi_i\right) + \frac{\partial}{\partial t}(\psi_f^*\psi_i) = 0 \tag{10}$$

This is a promising result for our purposes, since it has the form of an equation of continuity with i and f equally represented and it holds automatically for any two independent wavefunctions $\psi_i$ and $\psi_f$ that are both solutions of the Schrödinger equation.

Like (9), equation (10) implies a conserved quantity. This is easily demonstrated by performing an integral $d^3x$ over all space on each term in (10). Under the standard assumption that a wavefunction falls to zero as $\mathbf{x}$ goes to infinity, the integral of the first term in (10) is zero and we are left with:

$$\frac{\partial}{\partial t}\int_{-\infty}^{+\infty}\psi_f^*\,\psi_i\,d^3x = 0 \tag{11}$$

This result indicates that the quantity

$$a \equiv \int_{-\infty}^{+\infty}\psi_f^*(\mathbf{x},t)\,\psi_i(\mathbf{x},t)\,d^3x \tag{12}$$

is conserved through time. In particular, if it is non-zero at one instant of time, it must continue to be non-zero for other times.

Before using the new equation of continuity further, there is a need to rectify two minor complications which have arisen in the transition from (9) to (10). First, a normalization factor given by the amplitude a in (12) needs to be introduced to ensure that total probability remains equal to one. Second, the equation is no longer real, which will be avoided simply by taking the real part of it. This also makes it fully symmetric with respect to $\psi_i$ and $\psi_f$. Equation (10) thus becomes modified to the form[7]:

$$\nabla\cdot\mathrm{Re}\left(\frac{\hbar}{2ima}\psi_f^*\overleftrightarrow{\nabla}\psi_i\right)+\frac{\partial}{\partial t}\mathrm{Re}\left(\frac{1}{a}\psi_f^*\psi_i\right)=0 \qquad (13)$$

Note that we have been able to move the quantity a inside the time derivative in (13) because, from (11), it is independent of t.

Equation (13) has now been put into an appropriate form to provide expressions for a causally symmetric model. Specifically, comparing equations (6) and (13) points to the two identifications:

$$\rho(\mathbf{x},t)=\mathrm{Re}\left(\frac{1}{a}\psi_f^*\psi_i\right) \qquad (14)$$

and

$$\mathbf{v}(\mathbf{x},t)=\frac{\mathrm{Re}\left(\frac{\hbar}{2ima}\psi_f^*\overleftrightarrow{\nabla}\psi_i\right)}{\mathrm{Re}\left(\frac{1}{a}\psi_f^*\psi_i\right)} \qquad (15)$$

Equations (14) and (15) are the basis of the proposed causally symmetric version of Bohm's model. They contain $\psi_i$ and $\psi_f$ on an equal footing and should be compared with equations (1) and (2) of the original model[8].

The model also requires the following statistical assumption, which will be relevant in later sections: If the final wavefunction $\psi_f$ extending back from the future corresponds to one of the possible outcomes of a subsequent measurement[9], the conditional probability of $\psi_f$ given $\psi_i$ is:

$$\rho\left(\psi_f\middle|\psi_i\right)=|a|^2 \qquad (16)$$

where the amplitude a is defined in (12).

[7] At this point we can also consider the alternative of defining the normalizing factor to be the **real part** of expression (12). However, this would lead to a different result later in section 6(b).

[8] The physical interpretation of the case where the denominator of (15) is zero is stated in section 9 after the groundwork is laid in section 7.

[9] If a series of measurements is carried out, this assumption refers to the next measurement to be performed.

The obvious objection that can be made at this point is that the probability density (14) is not positive definite[10]. This will be dealt with in detail in section 7. Normally the prediction of a negative probability would be fatal for any proposed theory. It is a remarkable fact, however, that the introduction of backwards-in-time phenomena allows a natural interpretation of some negative probability expressions (in particular, those that are the basis of the new model). Therefore, unlike in other cases where negative probabilities have been mooted, there is no serious problem here. Indeed, one can argue that they should be expected in any truly causally symmetric model. In any case, at this point it simply needs to be stated that the model does not predict negative probabilities for the **outcomes of measurements**, so no meaning need be given here for such a notion. Indeed, using the word "probability" here may be a little misleading, but we will persist with it and leave the detailed explanation to section 7.

The discussion in the section 7 will be seen to be necessarily relativistic, whereas the considerations above have been limited to the non-relativistic case for simplicity and for ease of comparison with Bohm's original formalism. This situation will be rectified in section 14, where a causally symmetric model for the Dirac equation will be formulated.

## 5. Retrocausal influence on particle velocity

To demonstrate that the particle velocity defined by (15) really is retrocausally affected by future circumstances, consider two separate particles each having an identical initial wavefunction $\psi_i$ from time $t_1$ onwards. If we choose to perform measurements of different non-commuting observables on the particles at a later time $t_2$, they will have different final wavefunctions $\psi_f$ extending back from $t_2$ to $t_1$. In particular, these different $\psi_f$ 's will be eigenfunctions of the respective observables measured[11]. Since the velocity expression (15) is obviously dependent on $\psi_f$, it then follows that the velocity values at any intermediate time between $t_1$ and $t_2$ will be different for the two particles. Hence the type of measurement chosen at $t_2$ has a bearing on the physical reality existing at an earlier time, which constitutes retrocausality. This example also indicates the way in which our initial notion of retrocausality has been given a specific mathematical form.

Note further in this example that it is not possible to interpret the two $\psi_f$ 's as instead originating at some earlier time such as $t_1$, independent of the future measurements at $t_2$, and then propagating forwards in time. This is because these $\psi_f$ 's are eigenfunctions of two different observables that will subsequently be chosen freely[12] by the experimenter at $t_2$. It would be inexplicable why, for each particle, the $\psi_f$ that arises randomly at $t_1$ always happens to be an eigenfunction of the correct observable to be nominated and measured later at $t_2$. The only explanation is that each $\psi_f$ must be retrocausally determined by the choice at $t_2$.

---

[10] It is, of course, possible to devise alternative expressions for $\rho(\mathbf{x},t)$ that **are** positive. However, then the proposed probability distribution would not satisfy an equation of continuity and so probability would not be conserved, which is a more intractable problem.

[11] This will be demonstrated in section 13.

[12] An intuitive notion of free choice is being assumed here, although it is recognised that this is an area requiring further examination.

## 6. Some matters of consistency

It will be demonstrated briefly here that the probability expression (14) is quite consistent with what is observed when a measurement is actually performed. It should be kept in mind that the position probability distributions of Bohm-type models describe the position of a particle at all times, so that most of the times are **between** measurements. In terms of experimental agreement, it doesn't matter what is predicted there, since the distribution is hidden. We will now consider two simple cases to illustrate how (14) fits in with the usual quantum mechanical results.

**(a)** Consider a position measurement that gives a result $\mathbf{X}$ at some time T. Starting at earlier times t, the particle's final wavefunction must approach the form of a delta function:

$$\psi_f(\mathbf{x}) = \delta^3(\mathbf{x} - \mathbf{X}) \tag{17}$$

as T gets closer. Self-consistency of the model requires that the density $\rho(\mathbf{x},t)$ also becomes a delta function at T. To check that this is the case, it is more convenient to switch to Dirac bra-ket notation:

$$\begin{aligned}\rho(\mathbf{x},t) &= \mathrm{Re}(\tfrac{1}{a}\psi_f^*\psi_i) \\ &= \mathrm{Re}\frac{\langle\psi_f|\mathbf{x}\rangle\langle\mathbf{x}|\psi_i\rangle}{\langle\psi_f|\psi_i\rangle}\end{aligned} \tag{18}$$

Then, inserting the delta function (17), we have for time T:

$$\begin{aligned}\rho(\mathbf{x},T) &= \mathrm{Re}\frac{\delta^3(\mathbf{x}-\mathbf{X})\langle\mathbf{x}|\psi_i\rangle}{\langle\mathbf{X}|\psi_i\rangle} \\ &= \mathrm{Re}\frac{\delta^3(\mathbf{x}-\mathbf{X})\langle\mathbf{X}|\psi_i\rangle}{\langle\mathbf{X}|\psi_i\rangle} \\ &= \delta^3(\mathbf{x}-\mathbf{X}), \quad \text{as required.}\end{aligned} \tag{19}$$

So the distribution becomes positive at point $\mathbf{X}$ and zero everywhere else. This has the following consequence: Any negative value for the probability goes away as the time of the position observation is approached because the final wavefunction gradually dominates.

**(b)** It is also necessary to confirm that the causally symmetric expression (18) is consistent with the usual quantum mechanical distribution $|\psi_i(\mathbf{x})|^2$ predicted for position measurements. It needs to be pointed out here that expression (18) actually represents the **conditional** probability density given both the initial and final states:

$$\rho\left(\mathbf{x}\middle|\psi_i,\psi_f\right)=\mathrm{Re}\frac{\langle\psi_f|\mathbf{x}\rangle\langle\mathbf{x}|\psi_i\rangle}{\langle\psi_f|\psi_i\rangle} \tag{20}$$

Normally, however, the final state is not known and we need instead the conditional probability given the initial state alone. Obtaining the latter requires the use of our statistical assumption (16), which will be expressed in the more convenient form:

$$\rho\left(\psi_f\middle|\psi_i\right)=\left|\langle\psi_f|\psi_i\rangle\right|^2 \tag{21}$$

In anticipation of the analysis in section 13, we will assume here that the range of possible $\psi_f$ 's is restricted to the possible outcomes of the measurement that will subsequently be performed.

We proceed by employing the following general rule involving the joint probability distribution ρ(a,b) for two quantities a and b:

$$\rho(a,b)=\rho(a|b)\,\rho(b) \tag{22}$$

In our particular case this allows us to write:

$$\rho\left(\mathbf{x},\psi_f\middle|\psi_i\right)=\rho\left(\mathbf{x}\middle|\psi_i,\psi_f\right)\rho\left(\psi_f\middle|\psi_i\right) \tag{23}$$

in which the left hand side is conditional on $\psi_i$ alone. Inserting expressions (20) and (21) into the right hand side of (23) then gives:

$$\begin{aligned}\rho\left(\mathbf{x},\psi_f\middle|\psi_i\right)&=\mathrm{Re}\frac{\langle\psi_f|\mathbf{x}\rangle\langle\mathbf{x}|\psi_i\rangle}{\langle\psi_f|\psi_i\rangle}\left|\langle\psi_f|\psi_i\rangle\right|^2\\&=\mathrm{Re}\langle\psi_i|\psi_f\rangle\langle\psi_f|\mathbf{x}\rangle\langle\mathbf{x}|\psi_i\rangle\end{aligned} \tag{24}$$

Finally, summing over all possible final states $\psi_f$ under the assumption that these constitute a complete orthonormal set[13] yields the result:

$$\begin{aligned}\rho\left(\mathbf{x}\middle|\psi_i\right)&=\langle\psi_i|\mathbf{x}\rangle\langle\mathbf{x}|\psi_i\rangle\\&=\left|\psi_i(\mathbf{x})\right|^2\end{aligned} \tag{25}$$

The usual quantum mechanical expression has therefore been obtained, which provides the desired consistency. In particular, if a position measurement is about to be performed with the $\psi_f$ 's being its possible outcomes, it has been shown that the correct position distribution is in existence beforehand.

The results of this section illustrate some ways in which the causally symmetric model meshes neatly with standard quantum mechanics. The plausibility of this model, however, depends mainly on the conclusions of the next section.

---

[13] This follows from our assumption that the $\psi_f$ 's are the possible outcomes of a subsequent measurement.

## 7. Interpretation of negative probabilities

The aim of this section is to examine the usual formalism describing probability density for a particle's position and thereby understand the meaning of a negative value for this probability. The interpretation given below is not new, but previously it raised logical questions which now resolve themselves naturally in the context of retrocausality.

From here on we will adopt the convention of setting $\hbar = c = 1$. The relativistic formalism for probability current will now be briefly summarized. We return to the equation of continuity given earlier in equation (6):

$$\nabla \cdot (\rho \mathbf{v}) + \frac{\partial \rho}{\partial t} = 0 \tag{26}$$

which can be rewritten in relativistic notation as:

$$\partial_\nu (\rho_0 u^\nu) = 0 \tag{27}$$

where:

$\rho_0(\mathbf{x}, t)$ is the **rest** probability density, i.e., the probability density in the local rest frame of the probability flow at the space-time point $(\mathbf{x}, t)$,

$u^\nu = \dfrac{dx^\nu}{d\tau}$ is the 4-velocity of the flow at $(\mathbf{x}, t)$,

$\tau$ is the proper time taken along the 4-dimensional flow line at $(\mathbf{x}, t)$,

$x^\nu$ ($\nu = 0,1,2,3$) represents the coordinates t,x,y,z,

$\partial_\nu$ represents the partial derivative $\dfrac{\partial}{\partial x^\nu}$,

and a summation over $\nu$ is implied.

The quantity $\rho_0 u^\nu$ is known as the 4-current density and equation (27) states that its 4-divergence is zero. The rest density $\rho_0$ is an invariant, while $u^\nu$ is a 4-vector. Hence the current density $\rho_0 u^\nu$ is a 4-vector.

Now, comparing equations (26) and (27), the connection between the probability density $\rho$ and the rest density $\rho_0$ is identified to be:

$$\rho = \rho_0 u^0 \tag{28}$$

where $u^0$ is the time component of the 4-velocity. From this equation, the basic point is as follows:

> The probability density $\rho$ is seen to be **the time component of a 4-vector** in space-time. Hence the meaning of a negative value for this probability density at a particular point is simply that the time component of the current density is negative and so **the current density 4-vector is pointing backwards in time** at that point.

This is illustrated in the space-time diagram of Fig. 1. Only one of the three spatial components $\rho_0 u^i$ (i = 1,2,3) of the 4-current density can be shown since the diagram is 2-dimensional. The magnitude of this 4-vector is $\rho_0$, which will be relevant later.

To pursue this notion further, the flow line shown in Fig. 2 will be considered. Of course, such a line is generally viewed as not being physically permissible, but it will be useful as an example here. Now, since the current density 4-vector is directed backwards in time between space-time points 2 and 3, the probability density would be negative along this segment.

We can therefore draw the conclusion that negative probability would be a meaningful concept if probability flows such as that shown in Fig. 2 could occur in physics. Note that the rest density $\rho_0$ always remains positive. From equation (28), the density $\rho$ simply becomes negative when the 4-velocity component $u^0$ becomes negative. Recall that rest density is defined to be the density in the local rest frame of the flow. Such a rest frame can always be defined. It is straightforward to extend the concept of a reference frame to motions faster than light and paths backwards in time[14].

To consider further the likelihood of negative probabilities being relevant in physics, it will now be more convenient to focus directly on world lines of particles, rather than on lines of probability flow. To this end, let us tentatively examine the viability of the curve in Fig. 2 as a possible world line for a particle. Such a world line is generally viewed as being ruled out for several reasons. For example, (i) the particle behaves in a way that has never been observed, (ii) the particle goes faster than light, (iii) the particle goes backwards in time, (iv) the particle could be used to create causality paradoxes, (v) the particle passes smoothly through the "light barrier". Actually, however, none of these points constitutes a fatal objection here.

In response to (i), "doubling back" of the particle's world line only occurs at times **between measurements** and is therefore hidden. Detection of such behaviour would require a position measurement but, as discussed in section 6(a), the particle returns to normal behaviour as the time of the next position measurement approaches[15]. In response to (ii), many authors have pointed out that faster-than-light particles (tachyons) are consistent with special relativity. It is simply that they have never been observed experimentally. The response to (iii) is similar to that for (ii), since faster-than-light motion becomes backwards in time when viewed from an appropriately chosen, different frame of reference. In response to (iv), the particle's motion is beyond our control between measurements and so not able to be manipulated to create causality problems.

[14] For example, the time axis of the rest frame at point P in Fig. 2 is defined to be tangential to the flow line and in the direction of the arrowhead shown, while the spatial axes of the frame are defined to lie in the 3-dimensional hyperplane orthogonal to this time axis (orthogonality being well defined in Minkowskian geometry). Further details can be found in Sutherland and Shepanski (1986).

[15] The measurement clearly needs to have some retrocausal influence on the particle in order for this to occur. However, the considerations of section 6(a) indicate that such an evolution can happen quite naturally.

Finally, in response to point (v), there is no need for the classical laws of motion to remain valid in this quantum situation[16].

Since a world line that turns backwards in time cannot definitely be ruled out on theoretical grounds, it remains now to look at whether it might be a useful notion. Referring back to expression (14) for our position probability density, we are faced with the fact that this expression can only be explained in terms of continuous and smooth world lines if we are willing to permit world lines such as in Fig. 2. These are therefore being postulated here as being an essential (and perhaps a natural?) part of a causally symmetric model.

Before proceeding on, it should be mentioned that there is another possible interpretation that could be adopted for Fig. 2, namely that it simply represents the creation of a particle-antiparticle pair at point 3, followed by particle-antiparticle annihilation at point 2. This is certainly an equivalent way of viewing the situation, although such creation and annihilation events are normally represented with sharp vertices at 2 and 3 rather than smoothly curved ones, allowing compatibility with slower-than-light propagation. This alternative description involving particle-antiparticle pairs will not be employed here for three reasons. Firstly, the points 2 and 3 at which the world line reverses its time direction are actually both frame dependent, so that different observers will not agree in specifying the precise space-time event at which creation or annihilation occurs. Secondly, the single-particle perspective involves a single proper time variable $\tau$ increasing continuously along the world line as per the arrowheads shown, whereas the creation-annihilation view would require separate proper time variables for the three particles, changing discontinuously at the two (artificially generated) vertices. Thirdly, the proposed application of such paths is intended to be in quantum mechanical scenarios where they would not be directly observable anyway, so there is no need to think in terms of "what would actually be observed".

## 8. Overcoming a possible objection to negative probabilities

The particle-antiparticle viewpoint just discussed leads into another objection that should be considered. To introduce the argument, the world line we have been considering is presented again in Fig. 3, but with a particular time t highlighted for attention.

Suppose a position measurement of the x coordinate is performed at this time t. For simplicity it will be assumed that the measurement is certain to detect both the particle plus any possible particle-antiparticle pair that is present. Furthermore, it will be assumed that any entity detected is absorbed by the apparatus (e.g., a photographic plate) and prevented from proceeding further. Now, what will be the result of the measurement? One possible view is that particles will be detected at each of A and C, with an antiparticle being detected at B. Another view, however, is that the only thing that will be detected is a particle at A, since the "single world line" viewpoint entails that absorption at A will prevent the particle from ever reaching B and C.

[16] In any case, it is easy to construct a model which adheres to the usual classical laws yet accommodates passage through the light barrier. The issue to address is that special relativity does not permit a particle of non-zero rest mass to travel at the speed of light, since this would require infinite energy. To deal with this, one can simply add the assumption that the particle's rest mass varies appropriately with position (in a way dependent on the particle's wavefunction) so that it becomes zero at the instant when the particle passes through the light barrier. Such a model was outlined by de Broglie (1960, ch. 10) in a proposed relativistic extension of his hidden variable work. Of course, by definition, rest mass does not vary with velocity. However, there is nothing to prevent one postulating that rest mass varies with position.

Such an argument would raise questions for a model that involves world lines "doubling back" without involving retrocausality. The argument is easily avoided, however, once retrocausality is included as well. In the model we are considering here, the backwards-in-time effect of the position measurement (i.e., the influence of $\psi_f$) ensures the world line must straighten out as it approaches the measurement time, so that it arrives at only one point at time t. This is demonstrated trivially in section 6(a), where imposing the final boundary condition that the particle is detected at a certain location at time t results in a probability expression that is zero everywhere else at that time, so that it does not describe the presence of any other particle or antiparticle. The probability distribution may, of course, be spread out and containing negative regions at earlier times. As the measurement time approaches, however, the final wavefunction $\psi_f$ will dominate in expression (18) and the distribution will evolve gradually to a single point[17]. This discussion suggests that the notions of "doubling back" world lines and negative probabilities fit more harmoniously in combination with retrocausality than without it.

Having decided to pursue a causally symmetric model, one can actually adopt a more aggressive argument in favour of the possibility of world lines such as the one in Fig. 2. Models involving retrocausality arise most naturally from assuming the block universe picture, which in turn takes time and space to be similar. In such a context one can argue as follows: One would be surprised to find a particle whose world line can only ever point in the positive x direction, without ever doubling back in the negative x direction. But if time and space are on an equal footing, should we not be surprised if a world line can only point in the positive time direction without ever doubling back? Surely such a world line should be viewed as "unnatural"? Taking this attitude, negative probabilities for a particle's position are to be expected.

## 9. Some technical points

The sort of world lines we are considering is also reflected in the form of the 3-velocity expression (15):

$$\mathbf{v}(\mathbf{x},t) = \frac{\mathrm{Re}(\frac{\hbar}{2ima}\,\psi_f^* \overleftrightarrow{\nabla}\,\psi_i)}{\mathrm{Re}(\frac{1}{a}\,\psi_f^*\psi_i)}$$

This expression is infinite when its denominator $\mathrm{Re}(\frac{1}{a}\,\psi_f^*\psi_i)$ is zero, corresponding to points such as 2 and 3 in Fig. 2. This equation is not, however, able to indicate regions

[17] If the measurement is an imprecise one that only restricts the particle to a range of possible positions (e.g., by allowing the particle to pass through a slit of finite width), doubling-back behaviour such as in Fig. 3 may occur within the remaining range. This presents no conflict with experiment, however, since by definition the imprecise nature of the measurement prevents any such behaviour being detected.

where the world line has turned backwards in time. This information is provided by the time component of the 4-velocity, which is why it is more useful in this context to work in terms of a particle's 4-velocity rather than its 3-velocity.

The 4-velocity $u^{\nu} = \dfrac{dx^{\nu}}{d\tau}$ is defined in terms of the proper time $\tau$, which we are taking to be a variable that increases monotonically as we go along the world line from point 1 to point 4 in Fig. 2. It is clear that **$\tau$ needs to be always real**. This means that the usual definition:

$$\begin{aligned} d\tau &= \sqrt{dt^2 - dx^2 - dy^2 - dz^2} \\ &\equiv \sqrt{dx_{\mu} dx^{\mu}} \end{aligned} \tag{29}$$

which applies for a change $d\tau$ along a time-like segment of the world line, must be supplemented with the definition:

$$d\tau = \sqrt{-\,dx_{\mu} dx^{\mu}} \tag{30}$$

for the case of a space-like segment. This two-part definition for proper time is relativistically invariant because of the fact that **all observers agree as to whether any given 4-vector is time-like or space-like**. The definition can be written as:

$$d\tau = \begin{cases} \sqrt{dx_{\mu} dx^{\mu}} & \text{(time–like segment)} \\ \sqrt{-dx_{\mu} dx^{\mu}} & \text{(space–like segment)} \end{cases} \tag{31}$$

or, if preferred, summarized in the single expression[18]:

$$d\tau = \left| dx_{\mu} dx^{\mu} \right|^{\frac{1}{2}} \tag{32}$$

Finally, as can be seen from the 4-velocity relationship:

$$u_{\nu} u^{\nu} = \frac{dx_{\nu}}{d\tau} \frac{dx^{\nu}}{d\tau} \tag{33}$$

a consequence of (31) is that any time-like 4-velocity vector will satisfy the identity:

$$u_{\nu} u^{\nu} = 1 \tag{34}$$

whereas any space-like 4-velocity will satisfy:

$$u_{\nu} u^{\nu} = -1 \tag{35}$$

with the following identity holding for **any** 4-velocity vector:

$$\left| u_{\nu} u^{\nu} \right| = 1 \tag{36}$$

---

[18] Sutherland and Shepanski (1986)

This last result will be used in section 14.

## 10. Explanation of Bell Nonlocality

The aim of this section is to show how causal symmetry enables us to avoid the space-like effects and preferred frame needed in the standard Bohm model's description of the EPR/Bell experiment. The usual EPR/Bell arrangement is given in Fig. 4.

An initial state decays at event D into a pair of correlated particles and measurements are subsequently performed on the particles at $M_1$ and $M_2$, respectively. To simplify the discussion, we will take the $M_1$ measurement to occur earlier than the $M_2$ measurement, as shown in the diagram. The two measurements could be taken to be time-like separated if desired.

Before the first measurement is performed, the pair of particles is described by one overall wavefunction, which we will denote by $\psi_i(\mathbf{x}_1,\mathbf{x}_2)$. The two single-particle wavefunctions that subsequently arise from the measurements $M_1$ and $M_2$ will be denoted by $\psi_f(\mathbf{x}_1)$ and $\psi_f(\mathbf{x}_2)$, respectively. (To keep the notation simple, we are using the position coordinates to distinguish between the individual states.) We will now consider the standard description of the situation as events unfold.

**Standard Quantum Mechanical Description:**

Once the result of the measurement $M_1$ on the $1^{st}$ particle is known, the state of the other particle must be updated in order to make correct statistical predictions about the result of $M_2$. Specifically, the $2^{nd}$ particle must then be described by a single-particle wavefunction $\psi_i(\mathbf{x}_2)$ obtained from the scalar product of the $M_1$ outcome $\psi_f(\mathbf{x}_1)$ with the initial correlated state $\psi_i(\mathbf{x}_1,\mathbf{x}_2)$ via[19]:

$$\psi_i(\mathbf{x}_2)=\frac{1}{N}\int_{-\infty}^{+\infty}\psi_f^*(\mathbf{x}_1)\,\psi_i(\mathbf{x}_1,\mathbf{x}_2)\,d^3x_1 \qquad (37)$$

Hence the wavefunction description of the $2^{nd}$ particle changes as follows. At times between D and $M_1$ this particle is described by the wavefunction $\psi_i(\mathbf{x}_1,\mathbf{x}_2)$. Then, at times between $M_1$ and $M_2$, its appropriate wavefunction is $\psi_i(\mathbf{x}_2)$, as defined in (37). Finally, after $M_2$, the relevant wavefunction for the $2^{nd}$ particle is $\psi_f(\mathbf{x}_2)$. An analogous summary can be made of the successive wavefunctions of the $1^{st}$ particle.

We will now examine the further description given first by the standard Bohm model and then by the causally symmetric version in order to highlight the differences between these two models.

**Standard Bohm Model:**

In this case, the measurement $M_1$ exerts a space-like influence to cause a change in the $2^{nd}$ particle's trajectory compared with what it would otherwise have been. In accordance

---

[19] Here, N is a normalization constant ensuring $\int_{-\infty}^{+\infty}\psi_i^*(\mathbf{x}_2)\,\psi_i(\mathbf{x}_2)\,d^3x_2=1$

with Bell's theorem, this is necessary in order to allow for an effect on the $M_2$ measurement result.

In particular, at times between D and $M_1$, the 2nd particle's velocity is given by inserting the wavefunction $\psi_i(\mathbf{x}_1,\mathbf{x}_2)$ into equation (2) earlier to obtain:

$$\mathbf{v}_2(\mathbf{x}_1,\mathbf{x}_2)=\frac{\hbar}{2\mathrm{im}}\frac{\psi_i^*(\mathbf{x}_1,\mathbf{x}_2)\overleftrightarrow{\nabla}\psi_i(\mathbf{x}_1,\mathbf{x}_2)}{\psi_i^*(\mathbf{x}_1,\mathbf{x}_2)\psi_i(\mathbf{x}_1,\mathbf{x}_2)} \tag{38}$$

whereas, at times between $M_1$ and $M_2$, the relevant wavefunction is $\psi_i(\mathbf{x}_2)$ and so the 2nd particle's velocity is given by:

$$\mathbf{v}_2(\mathbf{x}_2)=\frac{\hbar}{2\mathrm{im}}\frac{\psi_i^*(\mathbf{x}_2)\overleftrightarrow{\nabla}\psi_i(\mathbf{x}_2)}{\psi_i^*(\mathbf{x}_2)\psi_i(\mathbf{x}_2)} \tag{39}$$

Note that, before the $M_1$ measurement, the velocities of the two particles are both calculated from the same wavefunction $\psi_i(\mathbf{x}_1,\mathbf{x}_2)$, which is defined in 6 dimensional configuration space. Therefore the 2nd particle's velocity depends on the 1st particle's position, which must be specified and inserted into (38). The situation at times after $M_1$ is different in that both particles have separate wavefunctions defined in 3 dimensions and so have independent velocity expressions.

**Causally Symmetric Model:**

In this case, we want to avoid any space-like influences between the particles. We know that the reduced wavefunction $\psi_i(\mathbf{x}_2)$ given by (37) is the correct one to use for predictions at the time of the measurement on the 2nd particle. Therefore, to avoid a space-like change, we need this wavefunction to be the correct one for determining the 2nd particle's velocity right back to the decay point D where the two particles separated, not just from $M_1$ onwards. The 2nd particle will thus be guided at all times between D and $M_2$ via a single-particle $\psi_i$ defined in 3 dimensions, even though this particle is initially part of a correlated pair. This possibility is available only in a causally symmetric theory, because the form of the wavefunction $\psi_i(\mathbf{x}_2)$ at times before the $M_1$ measurement depends on what type of measurement is subsequently chosen at $M_1$, which constitutes retrocausality.

Using the causally symmetric velocity expression (15) and inserting $\psi_i(\mathbf{x}_2)$ as the appropriate initial wavefunction, the specific form of the 2nd particle's velocity between D and $M_2$ is:

$$\mathbf{v}_2(\mathbf{x}_2)=\frac{\mathrm{Re}[\frac{\hbar}{2\mathrm{ima}}\psi_f^*(\mathbf{x}_2)\overleftrightarrow{\nabla}\psi_i(\mathbf{x}_2)]}{\mathrm{Re}[\frac{1}{\mathrm{a}}\psi_f^*(\mathbf{x}_2)\psi_i(\mathbf{x}_2)]} \tag{40}$$

where $\psi_f(\mathbf{x}_2)$ is the 2nd particle's final wavefunction and the amplitude a is given by:

$$a \equiv \int_{-\infty}^{+\infty} \psi_f^*(\mathbf{x}_2)\,\psi_i(\mathbf{x}_2)\,d^3x_2 \qquad (41)$$

The above scheme is in accordance with the space-time zigzag explanation of Bell's nonlocality suggested by a number of authors[20]. This explanation postulates the existence of a causal link along the path $M_1DM_2$ in Fig. 4. The type of measurement performed on the 1st particle at $M_1$ is assumed to have a bearing on that particle's state at earlier times, i.e., between $M_1$ and D. This in turn affects the other particle's state forwards in time from the decay point D, thereby affecting the result of the measurement at $M_2$. The apparent action at a distance in 3 dimensions then becomes a local connection when viewed from a 4-dimensional viewpoint.

In the present model, this general scheme has been given an explicit mathematical form. Recall that the causally symmetric model entails the 1st particle having (in addition to its initial wavefunction) a final wavefunction[21] $\psi_f(\mathbf{x}_1)$ evolving back from $M_1$ to D. Referring to Fig. 5, the arrowheads indicate the way in which the wavefunctions arising from the initial and left hand branches combine to produce a wavefunction for the right hand branch. In particular, the initial wavefunction $\psi_i(\mathbf{x}_1,\mathbf{x}_2)$ that arises from the decay of the original system[22] combines with the 1st particle's final wavefunction $\psi_f(\mathbf{x}_1)$ via the scalar product in (37) to give the 2nd particle's initial wavefunction $\psi_i(\mathbf{x}_2)$:

$$\psi_i(\mathbf{x}_2) = \frac{1}{N}\int_{-\infty}^{+\infty} \psi_f^*(\mathbf{x}_1)\,\psi_i(\mathbf{x}_1,\mathbf{x}_2)\,d^3x_1$$

## 11. Many-particle case: velocity

Consideration of the many-particle case has been postponed to this point to let the preceding discussion of Bell nonlocality dictate the way forward.

The equations in the previous section can now be generalized in a straightforward way to the case of n particles. Suppose we have a set of particles which have previously interacted and are therefore described by the configuration space wavefunction $\psi_i(\mathbf{x}_1,\ldots,\mathbf{x}_n)$. Suppose further that measurements are performed at time t on all particles except the jth one. Generalizing equation (37), the standard quantum mechanical description tells us that the jth particle should be described from time t onwards by the following 3-dimensional wavefunction:

---

[20] e.g., Costa de Beauregard (1953, 1977, 1987), Stapp (1975), Davidon (1976), Rietdijk (1978, 1987), Roberts (1978), Sutherland (1983, 1985, 1989, 1998), Price (1984, 1994, 1996), Cramer (1986), Hokkyo (1988), Miller (1996, 1997), Goldstein and Tumulka (2003).

[21] The tacit assumption that this final wavefunction is an eigenfunction of the $M_1$ measurement will be justified in section 13.

[22] Strictly speaking, it is the state of this original system, not $\psi_i(\mathbf{x}_1,\mathbf{x}_2)$, that should be written on the bottom arrowhead in Fig. 5. The two-particle wavefunction $\psi_i(\mathbf{x}_1,\mathbf{x}_2)$ simply encapsulates that aspect of the original state which is relevant thereafter, e.g., in determining the subsequent single-particle wavefunctions $\psi_i(\mathbf{x}_1)$ and $\psi_i(\mathbf{x}_2)$.

$$\psi_i(\mathbf{x}_j) = \frac{1}{N}\int_{-\infty}^{+\infty} \psi_f^*(\mathbf{x}_1)...\psi_f^*(\mathbf{x}_{j-1})\,\psi_f^*(\mathbf{x}_{j+1})...\psi_f^*(\mathbf{x}_n) \times \;\psi_i(\mathbf{x}_1,...,\mathbf{x}_j,...,\mathbf{x}_n)\, d^3x_1...d^3x_{j-1}\, d^3x_{j+1}...d^3x_n \tag{42}$$

where the various $\psi_f$ 's describe the measurement outcomes for the other particles.

Now, to avoid any space-like action at a distance when the particles are widely separated, we need the $j^{th}$ particle's velocity to depend on (42) **before** the measurements as well as after (rather than depending on $\psi_i(\mathbf{x}_1,...,\mathbf{x}_n)$ beforehand). Specifically, from expression (15), the $j^{th}$ particle's velocity must be given by:

$$\mathbf{v}_j(\mathbf{x}_j) = \frac{\mathrm{Re}[\frac{\hbar}{2ima}\,\psi_f^*(\mathbf{x}_j)\overleftrightarrow{\nabla}_j\,\psi_i(\mathbf{x}_j)]}{\mathrm{Re}[\frac{1}{a}\,\psi_f^*(\mathbf{x}_j)\psi_i(\mathbf{x}_j)]} \tag{43}$$

where $\psi_i(\mathbf{x}_j)$ is given by (42) and $\psi_f(\mathbf{x}_j)$ is this particle's final wavefunction. Hence, as in the two-particle case of the previous section, the velocity is defined in 3-dimensional space rather than configuration space.

A separate initial wavefunction similar to (42) can be introduced for each of the n correlated particles. Such 3-dimensional wavefunctions are easier to imagine as physically real than a wavefunction in 3n dimensions[23].

The above considerations assume that the system's **final** wavefunction is factorizable into single-particle wavefunctions (because we are assuming that n–1 measurements are performed). It is therefore not the most general case. To proceed further, we will insert (42) into (43) to write the $j^{th}$ velocity as:

$$\mathbf{v}_j(\mathbf{x}_j) = \frac{\mathrm{Re}[\frac{\hbar}{2ima}\int_{-\infty}^{+\infty}\psi_f^*(\mathbf{x}_1)\cdots\psi_f^*(\mathbf{x}_n)\overleftrightarrow{\nabla}_j\,\psi_i(\mathbf{x}_1,\cdots,\mathbf{x}_n)\,d^3x_1...d^3x_{j\text{-}1}\,d^3x_{j+1}...d^3x_n]}{\mathrm{Re}[\frac{1}{a}\int_{-\infty}^{+\infty}\psi_f^*(\mathbf{x}_1)\cdots\psi_f^*(\mathbf{x}_n)\,\psi_i(\mathbf{x}_1,\cdots,\mathbf{x}_n)\,d^3x_1...d^3x_{j\text{-}1}\,d^3x_{j+1}...d^3x_n]} \tag{44}$$

Now, the more general expression we are seeking must reduce to (44) when the final wavefunction of the system is factorizable. Hence the obvious generalization is:

$$\mathbf{v}_j(\mathbf{x}_j) = \frac{\mathrm{Re}[\frac{\hbar}{2ima}\int_{-\infty}^{+\infty}\psi_f^*(\mathbf{x}_1,\cdots,\mathbf{x}_n)\overleftrightarrow{\nabla}_j\,\psi_i(\mathbf{x}_1,\cdots,\mathbf{x}_n)\,d^3x_1...d^3x_{j\text{-}1}\,d^3x_{j+1}...d^3x_n]}{\mathrm{Re}[\frac{1}{a}\int_{-\infty}^{+\infty}\psi_f^*(\mathbf{x}_1,\cdots,\mathbf{x}_n)\,\psi_i(\mathbf{x}_1,\cdots,\mathbf{x}_n)\,d^3x_1...d^3x_{j\text{-}1}\,d^3x_{j+1}...d^3x_n]} \tag{45}$$

where a is given by $\langle\psi_f|\psi_i\rangle$ as usual. Since all the coordinates apart from $\mathbf{x}_j$ are integrated out, each particle's velocity continues to be expressible separately in 3-dimensional space for this general situation.

---

[23] It will be explained in the next section why the apparent loss of correlation associated with this formulation does not conflict with the predictions of quantum mechanics.

As discussed in section 3(ii), the $\psi_f$ of each particle will tend to become more correlated in going towards the past (i.e., the opposite of what occurs for each $\psi_i$ ), so the situation of the system having a factorizable final wavefunction, as described in equations (42) to (44), is expected to be a common one. Concerning the non-factorizable case of equation (45), however, it should be noted that a separate initial wavefunction in 3 dimensions cannot necessarily be assigned to each particle. By examining (42), one sees that the condition for the $j^{th}$ particle to be able to be assigned its own $\psi_i$ despite there being initial correlations is that this particle must have a separate final wavefunction:

$$\psi_f(\mathbf{x}_1,...,\mathbf{x}_n) = \psi_f(\mathbf{x}_1,...,\mathbf{x}_{j-1},\mathbf{x}_{j+1},...,\mathbf{x}_n)\,\psi_f(\mathbf{x}_j) \tag{46}$$

## 12. Many-particle case: probability density

It is straightforward to show that the general expression for velocity given above in equation (45) is consistent with an equation of continuity, in analogy to the discussion of section 4. Indeed, noting that the velocity in (45) has the usual form of current divided by density, the appropriate probability density that must be used in the continuity equation is simply the denominator of (45):

$$\rho\left(\mathbf{x}_j\middle|\psi_i,\psi_f\right) = \mathrm{Re}[\tfrac{1}{a}\int_{-\infty}^{+\infty}\psi_f^*(\mathbf{x}_1,\cdots\mathbf{x}_n)\,\psi_i(\mathbf{x}_1,\cdots,\mathbf{x}_n)\,d^3x_1...d^3x_{j-1}\,d^3x_{j+1}...d^3x_n] \tag{47}$$

This result provides the position probability distribution for the $j^{th}$ particle given both $\psi_i$ and $\psi_f$. In the likely event that $\psi_f$ is separable so that the $j^{th}$ particle has its own final wavefunction $\psi_f(\mathbf{x}_j)$ as shown in equation (46), the distribution (47) reduces to:

$$\rho\left(\mathbf{x}_j\middle|\psi_i,\psi_f\right) = \mathrm{Re}[\tfrac{1}{a}\psi_f^*(\mathbf{x}_j)\psi_i(\mathbf{x}_j)] \tag{48}$$

Here, $\psi_i(\mathbf{x}_j)$ is defined by analogy with (42) to be:

$$\psi_i(\mathbf{x}_j) = \int_{-\infty}^{+\infty}\psi_f^*(\mathbf{x}_1,...,\mathbf{x}_{j-1},\mathbf{x}_{j+1},...,\mathbf{x}_n)\,\psi_i(\mathbf{x}_1,...,\mathbf{x}_j,...,\mathbf{x}_n)\,d^3x_1...d^3x_{j-1}\,d^3x_{j+1}...d^3x_n \tag{49}$$

Returning to equation (47), note that it is a function of $\mathbf{x}_j$ alone because the other **x**'s are integrated out. Hence it provides a separate probability distribution for each of the n particles, instead of a single correlated distribution. One should contrast this with the usual many-particle expression of the standard Bohm model:

$$\rho\left(\mathbf{x}_1,\cdots,\mathbf{x}_n\middle|\psi_i\right) = \psi_i^*(\mathbf{x}_1,\cdots,\mathbf{x}_n)\,\psi_i(\mathbf{x}_1,\cdots,\mathbf{x}_n) \tag{50}$$

which is defined in 3n dimensional configuration space with the position probability density for the $j^{th}$ particle dependent on the positions of the other particles. This raises the question of whether the absence of correlations in the causally symmetric case is compatible with the predictions of quantum mechanics. It fact, however, the quantity $\mathbf{x}_j$ in equation (47) refers to the j[th] particle's position at times **between** measurements and is therefore not observable. To obtain the appropriate distribution for the outcome when a position measurement is performed on each of the n particles, it is necessary to return to our statistical assumption (21):

$$\rho\left(\psi_f \middle| \psi_i\right) = \left|\left\langle \psi_f \middle| \psi_i \right\rangle\right|^2 \tag{51}$$

and take the final wavefunction $\psi_f$ to consist of n position states arising from measurements performed at some final time T. These outcomes will be represented by upper case letters $\mathbf{X}_j$ to distinguish them from the hidden positions $\mathbf{x}_j$ (which relate to some earlier time t). Then from (51) we have trivially:

$$\rho\left(\mathbf{X}_1,\ldots,\mathbf{X}_n \middle| \psi_i\right) = \left|\left\langle \mathbf{X}_1,\ldots,\mathbf{X}_n \middle| \psi_i \right\rangle\right|^2 \tag{52}$$

which is equivalent to the standard distribution (50). The required correlations are therefore still present. Nevertheless, it is instructive to show that the hidden distribution (47) gradually becomes consistent with the observable one (52) as the time of the position measurement is approached and this is demonstrated in the Appendix.

Some further remarks are perhaps in order on this last point. It is well-known that, in the standard Bohm model, the probability distributions for most observable quantities (e.g., for a particle's momentum) do not necessarily conform to their predicted quantum mechanical forms at times other than measurement. Instead, the distribution being measured evolves into the correct form during the measurement interaction. This is shown to occur naturally in the Bohm theory of measurement and is obviously all that is required for agreement with experiment. The only exception is the position distribution, which conforms to the quantum mechanical prediction at all times in both the single and many-particle cases. In the causally symmetric model, by contrast, even the position distribution does not conform in general to the quantum mechanical expression at times other than measurement. The only exception this time is the single-particle case discussed earlier in section 6(b), where the usual position distribution is seen to be preserved.

## 13. Theory of measurement

A theory of measurement for the causally symmetric model will now be outlined[24]. Only a simplified treatment will be presented here, this being sufficient to allow further comparison between the two models.

Our main simplification will be to take the measuring apparatus as a macroscopic entity which can be treated classically so that its wavefunction need not be included explicitly in the argument. For example, the "apparatus" here may just take the form of a suitable potential inserted into the Schrödinger equation. In addition, any observable quantity discussed will be assumed to have a discrete spectrum of eigenvalues (the arguments being easily generalized to the case of a continuous spectrum).

---

[24] Some of the presentation in this section has been employed previously in Sutherland (1997, 1998).

We begin by describing some basic features of the theory of measurement associated with the standard Bohm model because these will continue to hold in the present case. Consider a particle with initial wavefunction $\psi_i$ . A measurement of an observable such momentum, energy, or spin is to be performed on this particle. An essential feature of any measurement is that it must allow us to distinguish between the different possible outcomes and identify the result. This means that the possible states of the observed system (or of something with which it interacts) must become **separated in space**. For instance, the separation may result from passing the wavefunction through a magnetic field, as in a Stern-Gerlach measurement of spin. This is a simple example of the type of measurement interaction we are assuming here. In such cases, the outcome of the separation stage of the measurement is that the initial wavefunction of the particle becomes a collection of spatially non-overlapping wave packets, the $j^{th}$ eigenfunction within the initial superposition becoming the $j^{th}$ packet. This takes place by continuous evolution via the relevant wave equation, e.g., the Schrödinger equation. As these wave packets gradually disunite, the particle (which is assumed to be travelling along a definite trajectory within the wavefunction) will flow continuously and smoothly into one of them. The measurement is completed by establishing in which packet the particle is located. This may be done in various ways. One may locate the particle by directly interacting with it, e.g., by blocking its path with a photographic plate. This identifies the correct eigenstate but has the disadvantage of immediately disrupting that state. Alternatively one may block all but one beam, so that particles in a particular, desired eigenstate continue on.

Note that, in order to obtain definite outcomes for experiments, standard quantum mechanics has to postulate that the wavefunction of a system spontaneously collapses to just one eigenstate upon measurement, even though such a postulate is not consistent with the basic axiom that wavefunctions evolve continuously as solutions of the Schrödinger equation. In Bohm's theory, by contrast, the definite outcome is obviously determined by the fact that the particle finishes up inside just one of the wave packets. Wavefunction collapse is then simply the decision to ignore the other packets in so far as they will have no further physical relevance.

The description so far is also applicable to the causally symmetric model. However, differences now arise through the introduction of the final wavefunction $\psi_f$. It needs to be kept in mind that the probability density (14) for the particle's position in the causally symmetric case involves of a product of initial and final wavefunctions:

$$\rho(\mathbf{x},t) = \mathrm{Re}(\tfrac{1}{a}\,\psi_f^*\psi_i)$$

and so the particle can only be found in regions where both these wavefunctions are non-zero, i.e., where they overlap. Now, in the discussion below, both wavefunctions will undergo branching into separate wave packets during measurement. Hence the analysis will involve looking at which branch of $\psi_i$ overlaps in space with $\psi_f$ , and vice versa. Furthermore, the conservation of probability described by equation (11) ensures that any overlap will persist through time. Specifically, if the particle is known to be present at some initial time, the Schrödinger equation requires $\psi_i$ and $\psi_f$ to evolve in such a way that there will continue to be some overlap throughout any subsequent series of measurements. Perhaps a useful way of viewing this process is that $\psi_i$ will spread forwards in time like separate fingers from a hand, whilst $\psi_f$ will spread into fingers backwards in time, with always at least one backwards and one forwards finger overlapping.

Now, in order to deduce how the particle's final wavefunction $\psi_f$ will behave during measurements, we will apply the following assumption of causal symmetry:

> The behaviour and properties of the final wavefunction are analogous to the initial wavefunction except for the time direction. This means one can always picture what the final wavefunction will do (in terms of branching, spreading, forming correlations, etc.) during measurements and other interactions by first thinking of how the initial wavefunction behaves and then just imagining the reverse.

With this in mind, the evolution of the usual wavefunction $\psi_i$ through measurement will be summarized briefly. The form of $\psi_i$ is initially arbitrary and not typically an eigenfunction of the measurement in question. (It may be an eigenfunction of some previous measurement, but that is not important here.) It then splits into spatially separated eigenfunctions of the relevant observable as it evolves through the region of the measurement interaction. If a series of measurements is performed, the wavefunction smoothly branches further with each measurement. The particle's trajectory always follows one branch only after each measurement and the other branches can then be deleted by choice as being irrelevant. The surviving wavefunction is therefore always a definite eigenfunction of the immediately preceding measurement. Finally, we emphasize here something that does **not** typically happen. The initial wavefunction does not approach the measurement region in the form of separate packets which merge for the first time during the measurement.

On this basis, the behaviour of the final wavefunction $\psi_f$ will be as follows: Its form will be arbitrary as it approaches the measurement time from the future. In particular, it will not typically be an eigenfunction of the observable in question. (It may be an eigenfunction of some measurement further in the future, but that is not important here.) Also, as it approaches the measurement region from the future, we can expect it will not typically consist of separate packets ready to merge for the first time[25]. As $\psi_f$ passes through the region of the measurement interaction towards our past, it will separate into spatially non-overlapping eigenstates of the observable concerned. The particle will be in one branch only and so the other wave packets can then be deleted as irrelevant. If a series of measurements is performed, the wavefunction smoothly branches further with each measurement as it evolves further towards our past. The resulting wave packets consist each time of eigenfunctions of the relevant observable. We can therefore conclude that, **when an observable quantity is measured, the form of $\psi_f$ before the measurement time will be a definite eigenfunction of that observable**[26].

---

[25] Hence in a Stern-Gerlach measurement on a spin-half particle, for example, we can expect that $\psi_f$ will be overlapping with only one of the two branches into which $\psi_i$ is split by the measurement.

There is a need here to verify that the $\psi_f$ eigenstate existing before the measurement is the same as the $\psi_i$ eigenstate continuing on after the measurement (i.e., that the $\psi_f$ eigenstate in question is actually in agreement with the measurement result). To this end, we will compare the evolution of both these wavefunctions though the measurement interaction towards our future. In the case of $\psi_i$, suppose it is the $j^{th}$ eigenfunction that, after spatial separation of the original superposition, subsequently contains the particle. In the case of $\psi_f$, on the other hand, we will suppose it is the $k^{th}$ eigenfunction that contains the particle before the measurement. This $\psi_f$ eigenfunction is present in the same region as the original superposition of $\psi_i$ eigenfunctions and is subject to the same Schrödinger evolution. Hence, in passing through the interaction time towards our future, it would evolve away in a different direction from the particle trajectory unless k is equal to j. Therefore we conclude that the $\psi_f$ state existing before the measurement is in fact the $j^{th}$ eigenstate and not some other one[27].

This concludes our discussion concerning the expected form and behaviour of $\psi_f$. The above considerations are sufficient for making further comparisons between the standard and causally symmetric Bohm models. The description given by standard quantum mechanics will also be contrasted.

First some features the two models have in common. Both models provide a continuous and smooth description of the measurement process, as opposed to the discontinuous wavefunction reduction of standard quantum theory. Also, both resolve the well-known measurement problem in the same way, namely through the choice of localized particles for their underlying ontology. This choice implies that each hidden trajectory must go into only one of the spatially separated eigenstates, thereby singling out a definite result. Both models also provide a Lorentz invariant description of wavefunction reduction in the single-particle case, whereas the instantaneous reduction assumed in standard quantum theory is not compatible with special relativity.

This brings us to our first point of difference. The causally symmetric model also introduces the possibility of Lorentz invariance in the many-particle case because it avoids the need for non-local effects.

Another difference between the two models relates to what boundary conditions determine a measurement outcome. In the standard Bohm model, given a particular wavefunction $\psi_i$, the outcome is determined by the particle's initial position, with different initial trajectories flowing into different post-measurement wave packets. In the causally symmetric model, by contrast, the decisive influence comes from the particular $\psi_f$ that is encountered. For example, if this $\psi_f$ overlaps only one of the branches arising from a given $\psi_i$, all possible trajectories will flow into that branch regardless of the particle's initial position[28].

---

[26] The word "before" refers here to **our** viewpoint. Of course, the resulting state will not necessarily still be an eigenfunction of the other observables measured in a series. Furthermore, the eigenfunction produced might not persist once the state undergoes time-evolution towards the past.

[27] Here we have implicitly employed the earlier assumption that $\psi_f$ is not expected to come back from later times as two or more branches that merge for the first time at the measurement.

Note that the two models give different reasons for why we cannot predict individual measurement outcomes in the quantum realm. In the standard Bohm case, to predict which packet the particle will enter would require knowledge about the particle's initial position within the wavefunction, which quantum mechanics does not permit us to gather. In the causally symmetric model, on the other hand, knowledge would be required in advance about $\psi_f$, i.e., about the future[29].

Finally, the standard Bohm model has the appealing feature that its theory of measurement allows the Born probability rule for any observable other than position to be deduced once the position distribution $|\psi_i(\mathbf{x})|^2$ is assumed. Such a derivation, however, is apparently not possible from the position distribution (14) of the causally symmetric model. Instead, the extra assumption (16) needs to be postulated to incorporate all probability expressions into the model.

## 14. Causally symmetric model for the Dirac equation

A relativistic version of the causally symmetric model developed above will now be formulated for the single-particle Dirac case. Taking $\hbar = c = 1$, the Dirac equation has the form:

$$\gamma^\mu \partial_\mu \psi + im\psi = 0 \tag{53}$$

and its hermitean conjugate is:

$$\partial_\mu \bar{\psi} \gamma^\mu - im\bar{\psi} = 0 \tag{54}$$

where:

$$\bar{\psi} = \psi^\dagger \gamma^0 \tag{55}$$

We proceed in the usual way to an equation of continuity. Multiplying (53) from the left by $\bar{\psi}$ and (54) from the right by $\psi$ and then adding yields the familiar result:

$$\partial_\nu (\bar{\psi} \gamma^\nu \psi) = 0 \tag{56}$$

---

[28]The initial position value will still play a role to the extent of determining precisely where the particle's trajectory is located within the surviving branch. However, this will obviously have no bearing on the measurement outcome itself.

[29] It is evident that, while the standard Bohm model is deterministic from the initial boundary conditions alone, this cannot be the case in any retrocausal model because the future is part of the cause, rather than just being the effect.

with the quantity in brackets being identifiable as the 4-current density. Now, it is easily seen that the derivation of equation (56) from (53) and (54) would remain valid if $\psi$ and $\overline{\psi}$ were two independent functions rather than being related as hermitean conjugates. We can therefore proceed as in the non-relativistic case of section 4 to modify (56) by (i) replacing $\psi$ with $\psi_i$ and $\overline{\psi}$ with $\overline{\psi}_f$, (ii) introducing a normalizing constant a, and (iii) taking the real part, to obtain the following equation of continuity:

$$\partial_\nu \operatorname{Re}(\tfrac{1}{a}\overline{\psi}_f \gamma^\nu \psi_i) = 0 \tag{57}$$

with:

$$a \equiv \int_{-\infty}^{+\infty} \overline{\psi}_f(\mathbf{x},t)\,\gamma^0 \psi_i(\mathbf{x},t)\,d^3x \tag{58}$$

Equation (57) will hold automatically provided $\psi_i$ and $\psi_f$ are solutions of the Dirac equation and so is suitable to serve as a probability-conserving starting point for a causally symmetric model. As mentioned in section 9, it will be more useful here to work in terms of the particle's 4-velocity $u^\nu = \dfrac{dx^\nu}{d\tau}$, rather than its 3-velocity $\mathbf{v} = \dfrac{d\mathbf{x}}{dt}$. Thus, comparing (57) with (27) points to the identification:

$$\rho_0 u^\nu = \operatorname{Re}(\tfrac{1}{a}\overline{\psi}_f \gamma^\nu \psi_i) \tag{59}$$

which provides a suitable current density expression for the model. In particular, the causally symmetric probability density for the particle's position is:

$$\rho \equiv \rho_0 u^0 = \operatorname{Re}(\tfrac{1}{a}\overline{\psi}_f \gamma^0 \psi_i) \tag{60}$$

Now, referring back to (36), we have the identity:

$$\rho_0 = \left|(\rho_0 u_\alpha)(\rho_0 u^\alpha)\right|^{\frac{1}{2}} \tag{61}$$

Hence, inserting (59) into (61), the rest density is found to be:

$$\rho_0 = \left|\operatorname{Re}(\tfrac{1}{a}\overline{\psi}_f \gamma_\alpha \psi_i)\operatorname{Re}(\tfrac{1}{a}\overline{\psi}_f \gamma^\alpha \psi_i)\right|^{\frac{1}{2}} \tag{62}$$

Combining this result with (59) then yields the following for the particle's 4-velocity:

$$u^\nu = \frac{\operatorname{Re}(\tfrac{1}{a}\overline{\psi}_f \gamma^\nu \psi_i)}{\rho_0} \tag{63}$$

where $\rho_0$ in the denominator is understood to be the expression in (62). Equations (60) and (63) are the basis of our causally symmetric Bohm model for the Dirac case.

In analogy to the non-relativistic model earlier, this relativistic formulation also needs the following statistical assumption:

If the final wavefunction $\psi_f$ extending back from the future is one of the possible outcomes of a subsequent measurement, the conditional probability of $\psi_f$ given $\psi_i$ is:

$$\rho\left(\psi_f \middle| \psi_i\right) = |a|^2 \tag{64}$$

where now the relevant amplitude a is defined by (58).

## 15. Conclusions

In this paper, a causally symmetric version of Bohm's model has been formulated. The aim has been for the advantages and disadvantages of such symmetry to be illustrated via a comparison of two otherwise similar models.

The advantages provided by causal symmetry are as follows. It reintroduces the possibility that the theory can be Lorentz invariant, with no need for a preferred reference frame at the hidden level. Also, the apparent non-locality highlighted by Bell's theorem can be given a local explanation from a 4-dimensional viewpoint. For the many-particle case, where the usual description is in terms of a single, correlated wavefunction defined in 3n-dimensional configuration space, causal symmetry allows each particle's velocity to be described by a separate expression in 3-dimensions. In addition, each particle can be described as being guided by its own 3-dimensional initial wavefunction $\psi_i(\mathbf{x})$, as long as the particle has a separate final wavefunction $\psi_f(\mathbf{x})$.

Causal symmetry also provides a viable physical meaning for the notion of negative probabilities. Finally, it even implies a possible reason for why tachyons are not observed directly and allows them to exist without causal loop problems.

On the other hand, some disadvantages are as follows. The causally symmetric model is not deterministic from the initial boundary conditions (although it becomes deterministic for predicting intermediate situations if the final boundary conditions are specified as well). Furthermore, although a logical explanation can be given for negative probabilities, this notion may nevertheless not appeal to everyone's taste. Finally, the equations of the causally symmetric version are not quite as simple as those of the original model.

Perhaps the main value to be gained from this formulation is that options which previously were suspected to be impossible (such as Lorentz invariance and 3-dimensional descriptions) are seen to be still in contention within a causally symmetric picture. Consequently, one should not lose heart in looking for an ontological model which can retain such features.

## Acknowledgements

The author would like to thank David Miller, Huw Price, Guido Bacciagaluppi, Sheldon Goldstein and Joseph Berkovitz for helpful comments on this manuscript.

## Appendix

This Appendix relates to section 12 and shows how the distribution (47) of the causally symmetric model becomes consistent with the usual quantum mechanical distribution (50) as the time of a position measurement is approached. The presentation is a generalization of the arguments in sections 6(a) and (b).

Starting with equation (47), the corresponding n-particle distribution at time t will have the following uncorrelated form:

$$\rho\left(\mathbf{x}_1,\cdots,\mathbf{x}_n,t\middle|\psi_i,\psi_f\right)=\prod_{j=1}^{n}\rho\left(\mathbf{x}_j,t\middle|\psi_i,\psi_f\right)$$

$$=\prod_{j=1}^{n}\mathrm{Re}[\tfrac{1}{a}\int_{-\infty}^{+\infty}\psi_f^*(\mathbf{x}_1,\cdots,\mathbf{x}_n,t)\,\psi_i\,\mathbf{x}_1,\cdots,\mathbf{x}_n,t)\,d^3x_1...d^3x_{j\text{-}1}\,d^3x_{j+1}...d^3x_n] \qquad \text{(A1)}$$

where the symbol $\Pi$ indicates a product. When the final wavefunction $\psi_f$ is a series of measured position states $\mathbf{X}_1,\cdots,\mathbf{X}_n$ at time T, the distribution (A1) can be expressed in bra-ket notation as follows:

$$\rho\left(\mathbf{x}_1,\cdots,\mathbf{x}_n,t\middle|\psi_i,\psi_f\right)$$

$$=\prod_{j=1}^{n}\mathrm{Re}\int_{-\infty}^{+\infty}\frac{\left\langle\mathbf{X}_1,\cdots,\mathbf{X}_n,T\middle|\mathbf{x}_1,\cdots,\mathbf{x}_n,t\right\rangle\left\langle\mathbf{x}_1,\cdots,\mathbf{x}_n,t\middle|\psi_i\right\rangle}{\left\langle\mathbf{X}_1,\cdots,\mathbf{X}_n,T\middle|\psi_i\right\rangle}\,d^3x_1...d^3x_{j\text{-}1}\,d^3x_{j+1}...d^3x_n \qquad \text{(A2)}$$

As the time of measurement gets closer (i.e., in the limit as t goes to T) this expression approaches the form:

$$\rho\left(\mathbf{x}_1,\cdots,\mathbf{x}_n,T\middle|\psi_i,\psi_f\right)$$

$$=\prod_{j=1}^{n}\mathrm{Re}\int_{-\infty}^{+\infty}\frac{\delta^3(\mathbf{x}_1-\mathbf{X}_1)...\delta^3(\mathbf{x}_n-\mathbf{X}_n)\left\langle\mathbf{x}_1,\cdots,\mathbf{x}_n,T\middle|\psi_i\right\rangle}{\left\langle\mathbf{X}_1,\cdots,\mathbf{X}_n,T\middle|\psi_i\right\rangle}\,d^3x_1...d^3x_{j\text{-}1}\,d^3x_{j+1}...d^3x_n$$

$$=\prod_{j=1}^{n}\mathrm{Re}\frac{\delta^3(\mathbf{x}_j-\mathbf{X}_j)\left\langle\mathbf{X}_1,\cdots,\mathbf{X}_n,T\middle|\psi_i\right\rangle}{\left\langle\mathbf{X}_1,\cdots,\mathbf{X}_n,T\middle|\psi_i\right\rangle}$$

$$=\prod_{j=1}^{n}\delta^3(\mathbf{x}_j-\mathbf{X}_j) \qquad \text{(A3)}$$

Now, equation (A3) provides the conditional probability density given both $\psi_i$ and $\psi_f$, whereas we want the result given $\psi_i$ alone because the final state is normally not known. To proceed, we construct a joint distribution for the particles to have hidden positions $\mathbf{x}_1,\cdots,\mathbf{x}_n$ <u>and</u> final state $\psi_f$ .This is achieved by using the general relationship (22) in the following form:

$$\rho\left(\mathbf{x}_1,\cdots,\mathbf{x}_n,\psi_f\middle|\psi_i\right)=\rho\left(\mathbf{x}_1,\cdots,\mathbf{x}_n\middle|\psi_i,\psi_f\right)\rho\left(\psi_f\middle|\psi_i\right) \qquad \text{(A4)}$$

then inserting (A3) and (21) to obtain for time T:

$$\rho\left(\mathbf{x}_1,\cdots,\mathbf{x}_n,\psi_f|\psi_i\right)=\left\{\prod_{j=1}^{n}\delta^3(\mathbf{x}_j-\mathbf{X}_j)\right\}\left|\langle\psi_f|\psi_i\rangle\right|^2 \quad \text{(A5)}$$

i.e.,

$$\rho\left(\mathbf{x}_1,\cdots,\mathbf{x}_n,\mathbf{X}_1,\cdots,\mathbf{X}_n|\psi_i\right)=\left\{\prod_{j=1}^{n}\delta^3(\mathbf{x}_j-\mathbf{X}_j)\right\}\left|\langle\mathbf{X}_1,\cdots,\mathbf{X}_n|\psi_i\rangle\right|^2 \quad \text{(A6)}$$

Finally, integration over the unknown final states $\mathbf{X}_1,\cdots,\mathbf{X}_n$ yields:

$$\rho\left(\mathbf{x}_1,\cdots,\mathbf{x}_n|\psi_i\right)=\left|\langle\mathbf{x}_1,\cdots,\mathbf{x}_n|\psi_i\rangle\right|^2 \quad \text{(A7)}$$

Hence the usual distribution emerges as the time of measurement is reached, as required.

A related point should be mentioned here. It is easily seen via a similar calculation (namely, inserting (A1) instead of (A3) into (A4)) that the probability density $\rho\left(\mathbf{x}_1,\cdots,\mathbf{x}_n|\psi_i\right)$ (i.e., conditional on $\psi_i$ alone) generally contains correlations at all times, but that the particular correlated form predicted by quantum mechanics only emerges as the measurement time is approached.

## References

Bohm, D. (1952a). A suggested interpretation of quantum theory in terms of “hidden variables” I. *Physical Review* ***85****, 166-179.*

Bohm, D. (1952b). A suggested interpretation of quantum theory in terms of “hidden variables” II. *Physical Review* ***85****, 180-193.*

Bohm, D. (1953). Comments on an article of Takabayasi concerning the formulation of quantum mechanics with classical pictures. *Progress of Theoretical Physics* ***9****, 273-287.*

de Broglie, L. (1960). *Nonlinear wave mechanics*. Elsevier, Amsterdam.

Costa de Beauregard, O. (1953). Méchanique quantique. *Comptes Rendus Académie des Sciences* ***236****, 1632.*

Costa de Beauregard, O. (1977). Time symmetry and the Einstein paradox. *Il Nuovo Cimento* ***42B****, 41-64.*

Costa de Beauregard, O. (1987). On the zigzagging causality EPR Model: Answer to Vigier and coworkers and to Sutherland. *Foundations of Physics* ***17****, 775-785.*

Cramer, J. G. (1986). The transactional interpretation of quantum mechanics. *Reviews of Modern Physics* ***58****, 647-687.*

Davidon, W. C. (1976). Quantum physics of single systems. *Il Nuovo Cimento* ***36B****, 34-39.*

Goldstein, S. and Tumulka, R. (2003). Opposite arrows of time can reconcile relativity and nonlocality. *Classical and Quantum Gravity* ***20****, 557-564.*

Hokkyo, N. (1988). Variational formulation of transactional and related interpretations of quantum mechanics. Foundations *of Physics Letters* ***1****, 293-299.*

Miller, D. J. (1996). Realism and time symmetry in quantum mechanics. *Physics Letters* ***A222****, 31-36.*

Miller, D. J. (1997). Conditional probabilities in quantum mechanics from a time-symmetric formulation. *Il Nuovo Cimento* ***112B****, 1577-1592.*

Price, H. (1984). The philosophy and physics of affecting the past. *Synthese* ***16****, 299-323.*

Price, H. (1994). A neglected route to realism about quantum mechanics. *Mind* ***103****, 303-336.*

Price, H. (1996). *Time's Arrow and Archimedes' Point*, Oxford University Press, Oxford.

Rietdijk, C. W. (1978). Proof of a retroactive influence. *Foundations of Physics* ***8****, 615-628.*

Rietdijk, C. W. (1987). Bell's theorem and retroactivity; on an objection by Sutherland. *Il Nuovo Cimento* ***97B****, 111-117.*

Roberts, K. V. (1978). An objective interpretation of Lagrangian quantum mechanics. *Proceedings of the Royal Society of London* ***A360****, 135-160.*

Stapp, H. P. (1975). Bell's theorem and world process. *Il Nuovo Cimento* ***29B****, 270-276.*

Sutherland, R. I. (1983). Bell's theorem and backwards-in-time causality. *International Journal of Theoretical Physics* ***22****, 377-384.*

Sutherland, R. I. (1985). A corollary to Bell's theorem. *Il Nuovo Cimento* ***88B****, 114-118.*

Sutherland, R. I. (1989). Implications of a causality paradox related to Bell's theorem. *Il Nuovo Cimento* ***104B****, 29-33.*

Sutherland, R. I. (1997). Phase space generalization of the de Broglie-Bohm model. *Foundations of Physics* ***27****, 845-863.*

Sutherland, R. I. (1998). Density formalism for quantum theory. *Foundations of Physics* ***28****, 1157-1190.*

Sutherland, R. I. and Shepanski, J. R. (1986). Superluminal reference frames and generalized Lorentz transformations. *Physical Review* ***D33****, 2896-2902.*

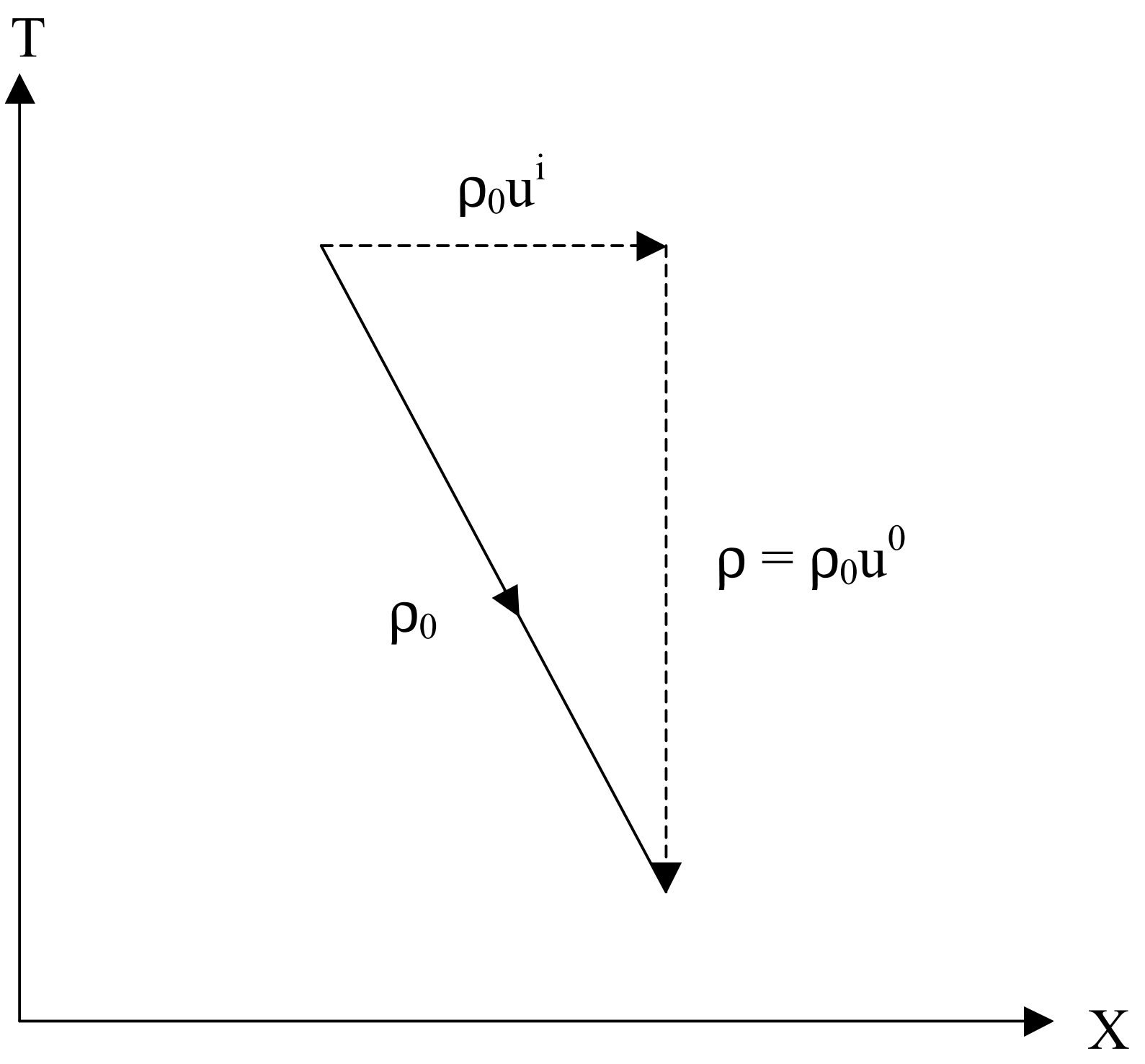

**Fig. 1**

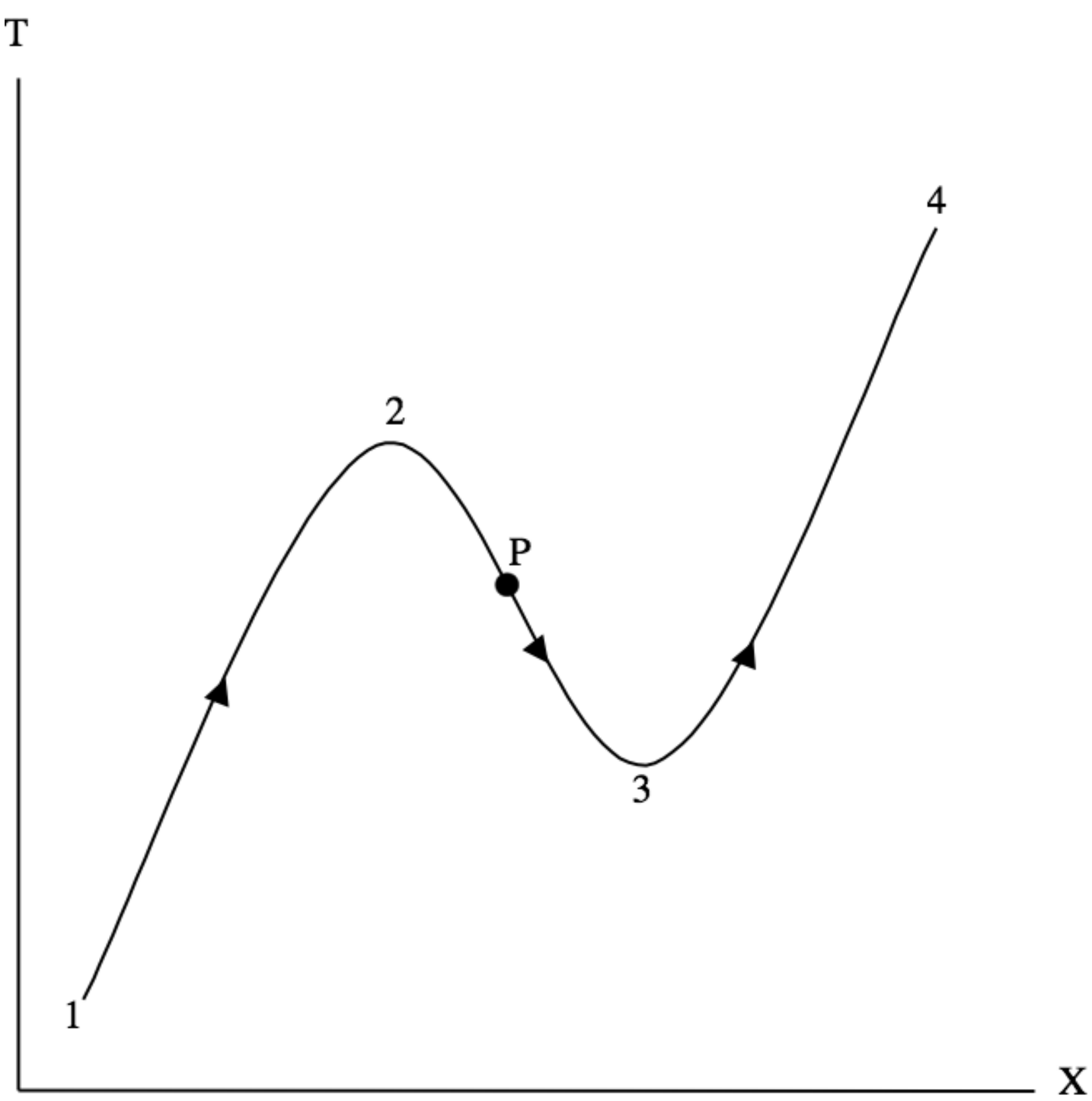

**Fig. 2**

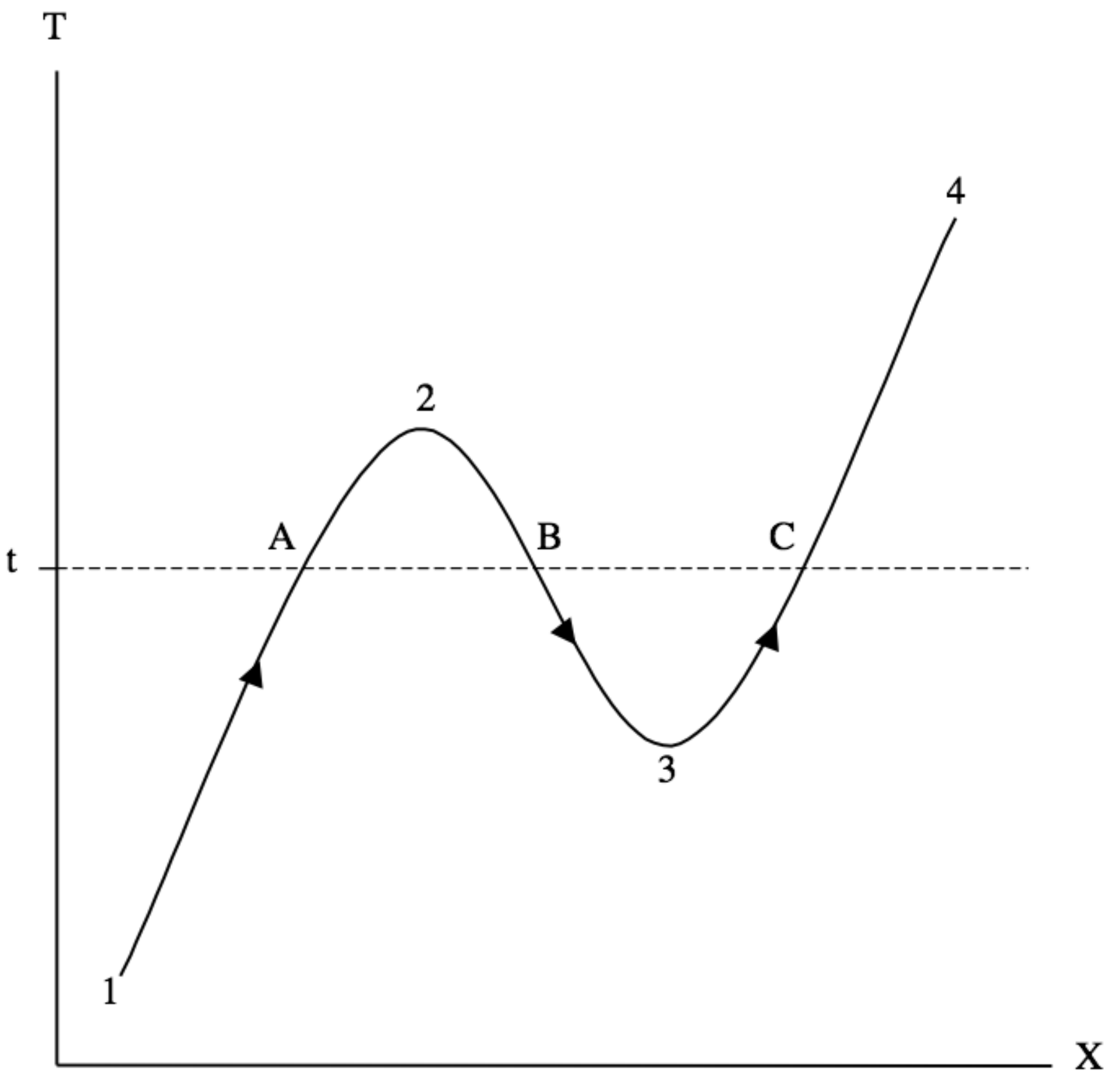

**Fig. 3**

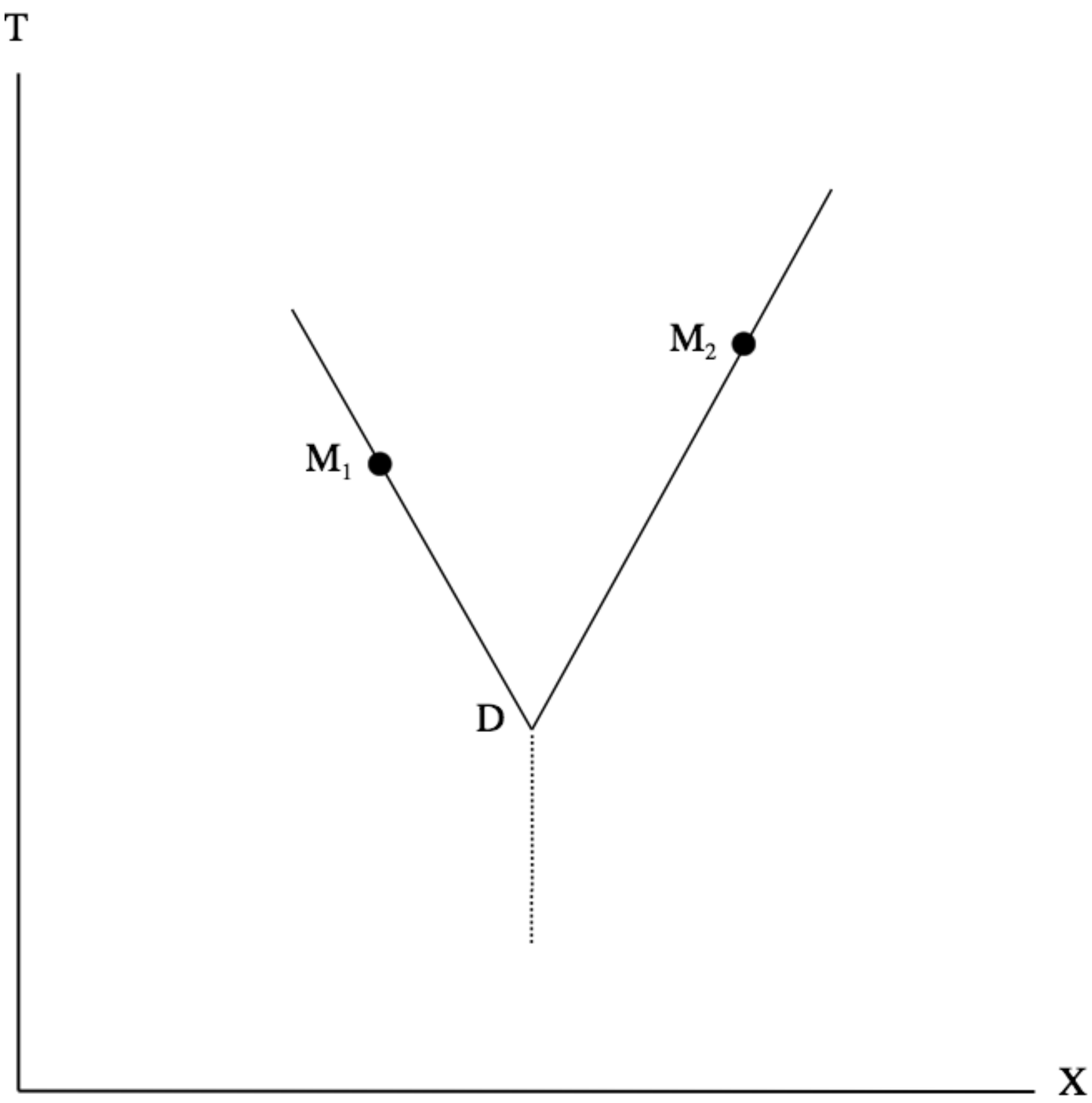

**Fig. 4**

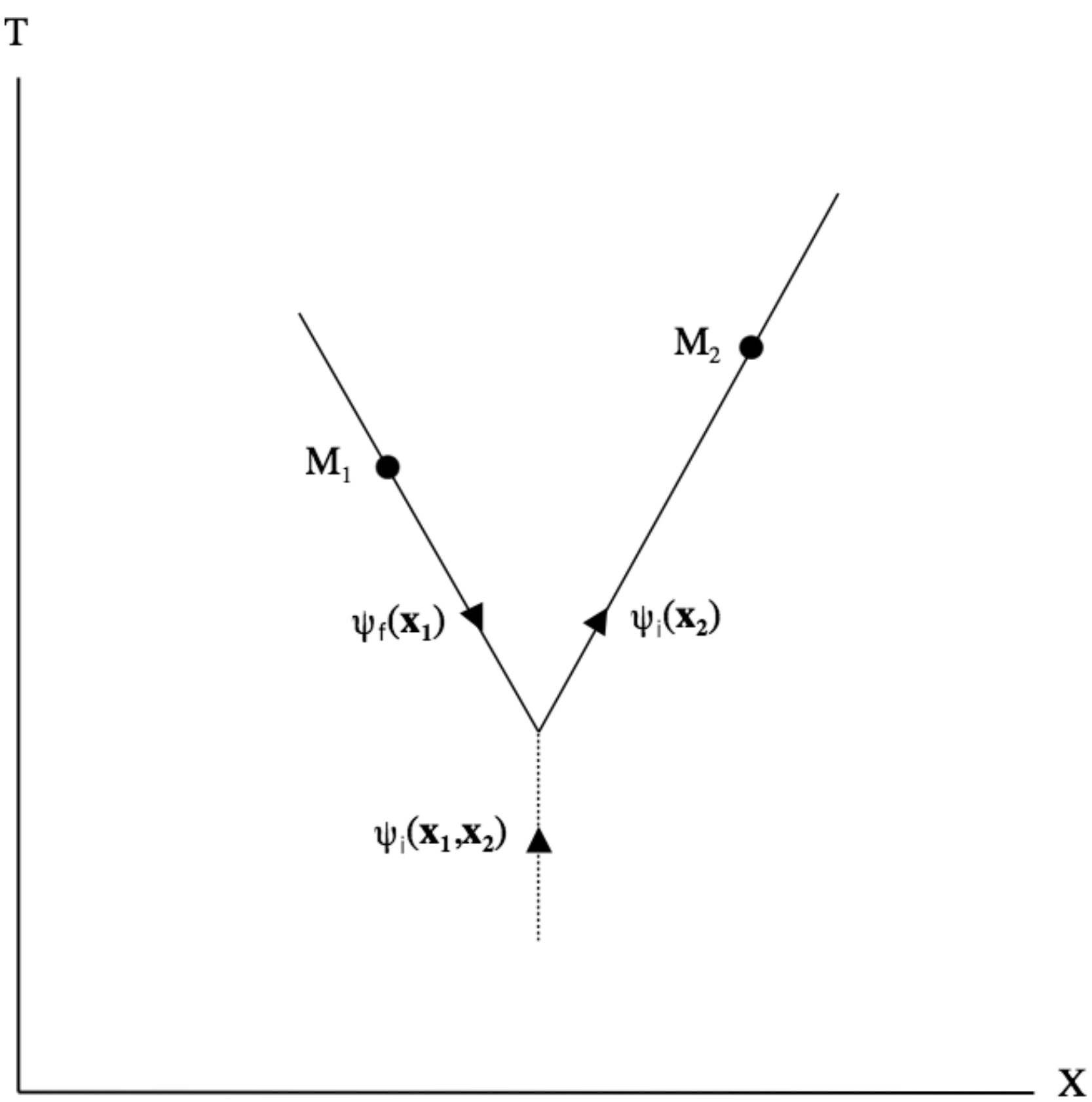

**Fig. 5**